# MONTE CARLO STUDY OF A WIDE RANGE OF MOLECULAR SPINTRONICS DEVICES

Pawan Tyagi[1], Christopher D'Angelo[2], Collin Baker

[1]Mechanical Engineering, School of Engineering and Applied Science, University of the District of Columbia, Washington, DC, 20008, USA, Email:ptyagi@udc.edu

[2]Department of Mathematics and Statistics, University of the District of Columbia, Washington DC 20008, USA

**ABSTRACT**

Molecular spintronics devices (MSDs) are highly promising candidates for enabling quantum computation and revolutionizing computer logic and memory. An advanced MSD will require the placement of magnetic molecules between the two ferromagnetic (FM) electrodes. Recent experimental studies showed that some magnetic molecules produced unprecedented strong exchange couplings between the two FM electrodes leading to intriguing magnetic and transport properties in a MSD. Future development of MSDs will critically depend on obtaining an in-depth understanding of the molecule induced exchange coupling, and its impact on switchability, functional temperature range, and stability. However, the large size of MSD systems and fragile device fabrication scheme continue to limit the theoretical and experimental studies of magnetic attributes produced by molecules in a MSD. This paper theoretically studies the MSD by performing Monte Carlo simulations (MCS). Our MCS encompasses the full range of MSDs that can be realized by establishing different kinds of magnetic interaction between magnetic molecules and FM electrodes. Our MSDs are represented by a 2D Ising model. We studied the effect of a wide range of molecule-FM electrode couplings on the basic properties of MSDs. This wide range covered (i) molecule possessing ferromagnetic coupling with both FM electrodes, (ii) molecule possessing antiferromagnetic coupling with both FM electrodes, and (iii) molecule possessing ferromagnetic coupling with one electrode and antiferromagnetic coupling with another FM electrode. Our MCS will enable the fundamental understanding and designing of a wide range of novel MSDs utilizing a variety of molecules and FM electrodes; these studies will also benefits nanomaterials based spintronics devices employing nanoclusters and quantum dots as the device elements.

**INTRODUCTION**

Molecular spintronics devices (MSDs) have attracted worldwide attention due to their potential to revolutionize logic and memory devices [1, 3]. A typical MSD is comprised of two ferromagnetic (FM) electrodes, coupled by molecular channels. Molecular channels with a net spin state are the basis of a large number of intriguing studies [9], which were either observed experimentally [4, 7] or were calculated theoretically [1]. Porphyrins [5], single molecular magnets [1] and magnetic molecular clusters [6] possess a net spin state and can be synthetically tailored to be employed in a MSD. Single molecular magnet-based MSDs have been widely discussed as the practical architecture for quantum computation [3]. However, real application of MSDs is impeded by the ongoing experimental difficulties in the key fabrication approaches [11, 13]. Despite a sluggish progress in realizing a commercially viable MSD, a number of experimental studies have shown that a magnetic molecule(s) between two FM electrodes may produce dramatic attributes on the overall MSD [9, 14]. However, a deeper understanding of the effect of molecular device elements on the magnetic attributes of MSDs cannot be studied experimentally with the popular MSD approaches. These approaches are largely based on using metal break junction and sandwiching molecule(s) between two FM electrodes [12]. The fundamental reason behind this persisting knowledge gap is that it is extremely challenging to study FM electrodes, at a gap of nm, before and after the creation of molecular device bridges between them. To circumvent experimental difficulties, a number of theoretical studies have been attempted to map the influence of magnetic interaction between molecules and FM electrodes of a MSD. Recently, a few DFT studies have started focusing on the interaction between magnetic molecules and *one FM film* [2]. These types of theoretical calculations produced an incomplete understanding and have the following limitations: (a) the DFT simulation cell contained only a few hundred restricted atoms and hence, they do not represent any realistic mass producible MSD; (b) to make the computation manageable numerous assumptions and approximations are employed, and; (c) the molecule interaction is only considered with one FM electrode [2]; in a real MSD molecule has to be connected with at least two electrodes. At present, there is no systematic method to understand the magnetic molecule induced magnetic characteristics on a MSD. The switching of device states and the operating temperature limit depends on the molecule induced magnetic properties of a MSD. This paper investigates the impact of tunable molecular device element on the MSD's magnetic properties by performing Monte Carlo Simulations (MCS).

## METHODOLGY

MCS are conducted to theoretically understand the effect of molecular device element in yielding the resultant magnetic properties on a MSD. To do so, we developed the MCS codes in C++ programming language. MCS have been successfully employed to study ferromagnetic systems using Ising models [8]. This highly mature simulation approach has a large number of algorithms and techniques to reveal a wealth of insights about the MSD systems. This study focuses on a 2D Ising model representing a MSD configuration, where molecules are placed between two FM electrodes (Figure 1A). To deal with the complex interaction between molecule and FM electrodes our MCS use the exchange interactions between molecules and the FM electrodes as the *tunable* parameters (Figure 1B). This strategy has two advantages: (i) there is no need to focus on the tedious calculation of determining exchange interactions between magnetic molecules and the FM electrodes [2], and (ii) MCS have the ability to study the effect of a wide range of molecules' spins without delving into molecule specific details. The key parameters included in this study are the exchange-coupling strengths of magnetic molecule with the two FM electrodes, thermal energy (system temperature ($T$) times Boltzmann constant ($k$)= $kT$), Heisenberg coupling strengths for the FM electrodes, dimensionality, and system size. The key observables in MCS are magnetization ($M$), heat capacity ($c$), and magnetic susceptibility ($x$). We critically investigated the phase transition points in a MSD.

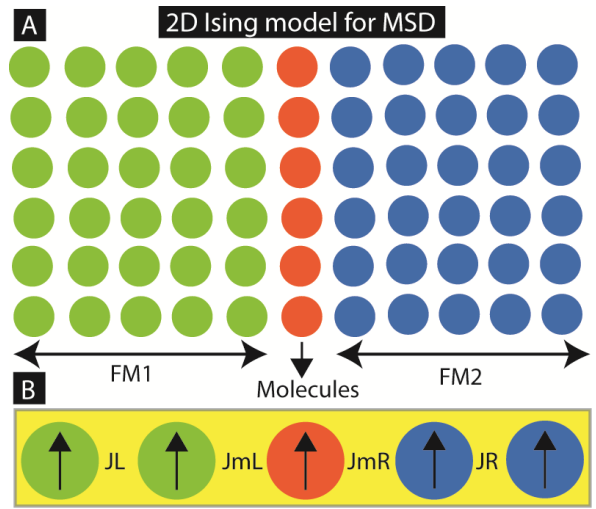

**Figure 1.** The MSD configuration for MCS: (a) 2D MSD model with simplified molecular representation in between two FM electrodes, (b) exchange coupling parameters to be used.

The MCS used the 2D Ising model and the Metropolis algorithm. The 2D version of the following energy term was utilized to govern the stable state of the system (Figure 1B).

$$E = -JL(\sum_{i \in L'} S_i S_{i+1} + \sum_{i \in L} S_i S_{i+w}) - JR(\sum_{i \in R'} S_i S_{i+1} + \sum_{i \in R} S_i S_{i+w}) - JmL(\sum_{i \in L-L'} S_i S_{i+1}) - JmR(\sum_{i \in R-R'} S_{i-1} S_i) \quad \text{(eq.1)}$$

In the above equation: *JL*, *JR*, *JmL*, and *JmR* are the Heisenberg exchange coupling strengths for the FM electrodes on the left, the right, between the left FM electrode and the molecule, and between the right FM electrode and the molecule, respectively (Figure 1). $S_i$ and $S_{i\pm1}$, $S_{i\pm w}$ are the spins of the nearest neighbors, with *w* being the width of the system. L and R are the sets of atoms in the left and right FMs, respectively, and their "prime" counterparts are subsets excluding the column nearest to the molecule, i.e. excluding the columns affected by the exchange couplings *JmL* and *JmR*. Initially, only the Heisenberg interaction was considered and a periodic boundary condition was employed [8]. Usage of the periodic boundary condition ensure that the spins on the one edge of the Ising lattice are the nearest neighbor to the corresponding spins on the opposite edge [8]. After choosing appropriate values for the Heisenberg exchange coupling coefficient, thermal energy (*kT*), and random spin states, a Markov process was set up to generate a new state. Under the Metropolis algorithm, the spin of a randomly selected site was flipped to produce a new state. New states were rejected if difference between the final and new energy (*ΔE*) satisfy both:

$$\Delta E > 0, \ e^{\frac{-\Delta E}{kT}} \leq r$$

where *r* is a uniformly distributed random variable in the half-closed interval [0,1). Here, ΔE is derived from equation (1) and denote the difference between the energy of the system before and after flipping spin of a randomly selected atom or molecules. The magnetization of the individual FM atoms and molecules is represented by *m*.

The *kT* is the measure of thermal energy of the 2D Ising model and has the same unit as exchange coupling parameters. To keep discussion generic and the exchange coupling parameters and *kT* are referred as the unitless parameters throughout this study.

New configurations were generated and observables were calculated. To determine the optimum point for the estimation of observables, after performing stability checks, magnetization for the overall MSD and the FM electrodes were recorded over *N* steps.

To investigate the effect of molecular exchange couplers on the properties of FM electrodes the effect of various parameters were studied on the magnetization (*M*), specific heat (*c*), and magnetic susceptibility (*x*) of the overall device. Simultaneously, *M*, *c*, and *x* for the left and right FM electrodes were also studied. These quantities for the left and right FM electrodes are denoted with *L* and *R* suffixes, respectively. Following mathematical expressions were programmed to calculate *c* and *x* in our Monte Carlo studies.

$$c = \frac{(kT)^2}{N}(<E^2> - <E>^2) \quad \text{(equation 2)}$$

$$x = (kTN)(<m^2> - <m>^2) \quad \text{(equation 3)}$$

This study mainly focused on the following three main cases under which molecular device elements possessed (i) ferromagnetic coupling with the both FM electrodes, denoted by BFMC, (ii) antiferromagnetic coupling with the both electrodes, denoted by BAFMC, and (iii) ferromagnetic coupling with one electrode and antiferromagnetic coupling with the another one, denoted by FMAFMC. The MSD with these situations are denoted by BFMC-MSD, BAFMC-MSD, and FMAFMC-MSD. For these three cases the effect of *kT* and the strength of molecular exchange coupling were studied. Most of the studies are performed with the 11 column and 10 rows (2D11x10) systems. In the 11x10 grid, a column of molecules is placed at $6^{th}$ column, so that on either side of the molecular column there will be 5x10 atoms. This convention is maintained for all the systems studied in our work. We studied other 2D system sizes and found the results consistent with 2D11x10 systems.

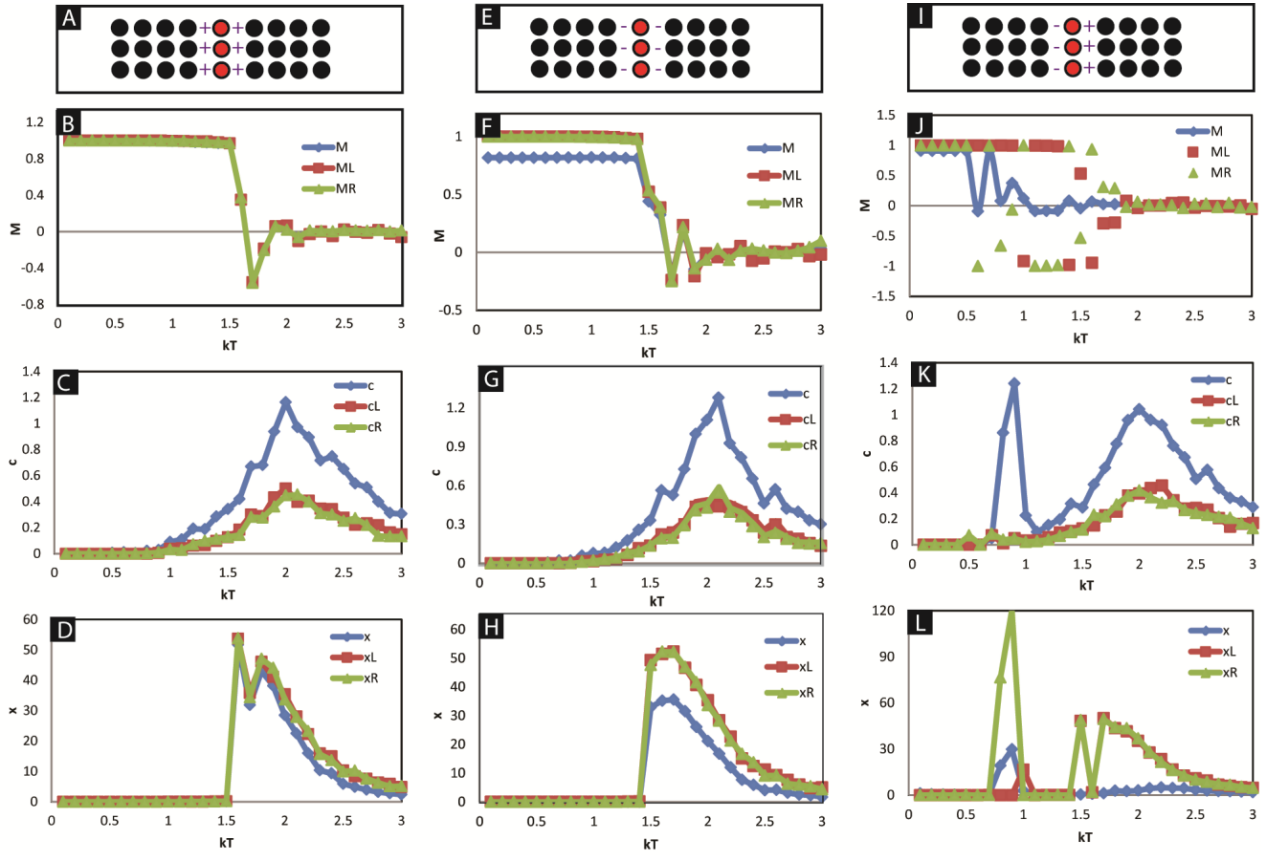

**Figure 2:** Effect of *kT* on magnetization (*M*) specific heat (*c*) and magnetic susceptibility (*x*) of 2D11x10 MSDs: (A)BFMC-MSD configuration and the effect of *kT* on its (B)*M*, *ML* and *MR*, (C) *c cL*, and *cR*, (D) *x*, *xL*, and *xR*. (E)BAFMC-MSD configuration and the effect of *kT* on its (F)*M*, *ML* and *MR*, (G) *c cL*, and *cR*, and (H) *x*, *xL*, and *xR*. (I)FMAFMC-MSD configuration and the effect of kT on its (J)*M*, *ML* and *MR*, (K) *c*, *cL*, and *cR*, and (L) *x*, *xL*, and *xR*.

In order to determine the duration of simulation after which a typical system reach in the equilibrium state stability test was performed. Magnetization data was recorded with the increasing number of iterations. In the beginning, spins on the atoms of the ferromagnetic electrodes and the molecules were assigned randomly or fixed in the same direction. Such stability tests were performed for all the MSD system sizes studied during this research. After reaching in stable state a number of measurements were made to calculate the observables. To keep the measurements uncorrelated time interval between two consecutive measurements of observables was determined by the dedicated simulations of correlation time [8].

**RESULTS AND DISCUSSION**

We began our studies with 2D 11x10 dimension of BFMC-MSD (Figure 2A). Initially, we studied the effect of thermal energy ($kT$) for $JL=JR=JmL=JmR=1$. In this case molecules produced no striking difference in magnetic properties as compared to a 2D11x10 ferromagnet (without any molecule). Magnetization($M$) versus $kT$ graph exhibited one major transition around $kT=1.5$. After this transition, overall magnetization became nearly zero (Figure 2B). Heat capacity ($c$) versus $kT$ graph provided additional insights about the phase transition (Figure 2C). For the BFMC case $c$ started changing around $kT=1$ and reached to its maximum value around $kT=2.0$. It is apparent that phase transition is completed around $kT=2.0$. Change in heat capacity is known as a major indicator of the phase change [8]. The magnetic susceptibility($x$) vs. $kT$ indicated a major change around $kT=1.5$ (Figure 2D), which is consistent with the $M$ vs. $kT$ data presented in Figure 2B.

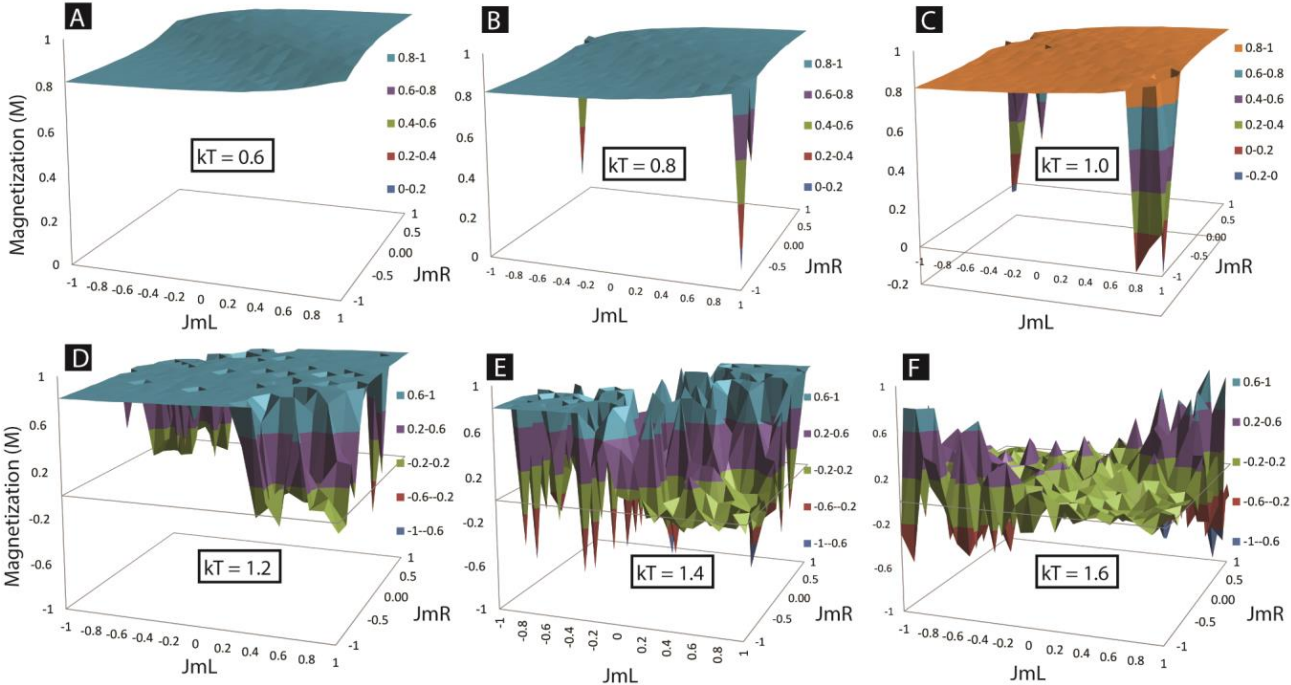

Figure 3: Effect of *JmL* and *JmR* on *M* at $kT$ = (A) 0.6, (B) 0.8, (C) 1.0, (D) 1.2, (E) 1.4, and (F) 1.6.

For the BAFMC-MSD similar observations were observed (Figure 2E-H). However, the total magnetization was smaller than that observed for the BFMC-MSD case (Figure 2F). This is due to the fact that the molecular column maintained magnetization direction in the opposite direction with respect to to two FM electrodes (Figure 2E). Heat capacity (Figure 2G) and magnetic susceptibility (Figure 2H) resembled with the single FM electrode and BFMC-MSD case.

The FMAFMC-MSD case produced quite intriguing results (Figure 2I-L). The *M* vs. *kT* graph showed three distinctive regions (Figure 2J): (i) low *kT* region where two FM electrodes and overall magnetization of the MSD was aligned in the same direction, (ii) medium *kT* region, between ~0.6 and 1.5, where two FM electrodes were aligned in the *opposite* direction and overall magnetization became nearly zero, (iii) high *kT* region after ~1.5 where magnetizations of the two electrodes and the overall device is close to zero. This third region is consistent with the appearance of highly disordered state in BFMC and BAFMC cases as well (Figure 2D and E). The *c* vs. *kT* graph (Figure 2J) confirms the appearance two major phase transitions: the first one around *kT*=0.8, due to the presence of molecules, and the second one completing around *kT* = 2.0, due to high thermal energy. The *x* vs. *kT* graph supports the observation of two-phase transitions (Figure 2L). According to both, *x* and *c* data the first transition was quite sharp as compared to the second transition.

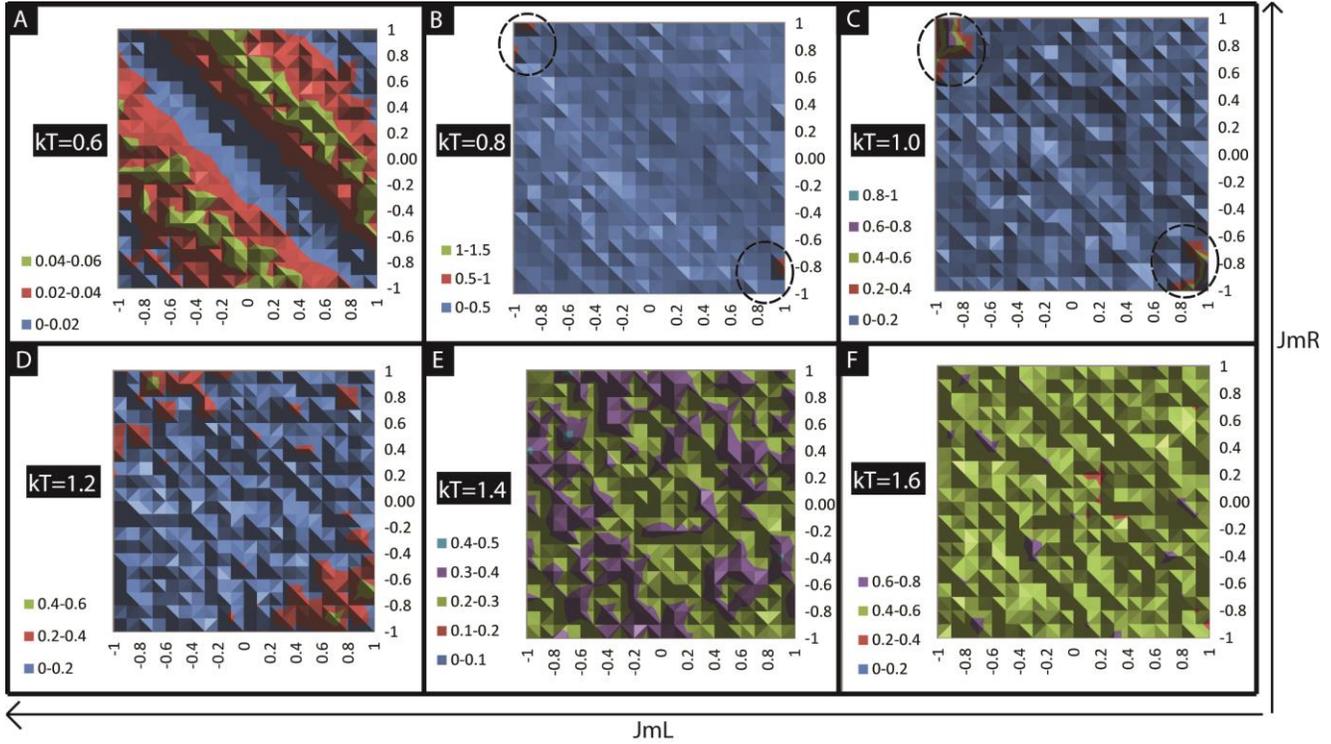

Figure 4: Effect of *JmL* and *JmR* on *c* at *kT* = (A) 0.6, (B) 0.8, (C) 1.0, (D) 1.2, (E) 1.4, and (F) 1.6.

Strength of molecular exchange coupling is expected to be the most important variable in determining the overall MSD properties. To study all the three cases, BFMC-MSD, BAFMC-MSD, and FMAFMC-MSD, (while studying the effect of *JmL* and *JmR*) their magnitude were changed from 1 to -1 at different *kT*. Different values of *kT* for this study were selected according to the phase transition stages in Figure 2. M remained unchanged for *kT*<0.8 (Figure 3A). However, for the negative values of both *JmL* and *JmR* (representing BAFMC case) the *M* was slightly lower than that for the positive values of both *JmL* and *JmR* (representing BFMC case). For *kT*=0.8, the MSDs possessing opposite sign of *JmL* and *JmR* with high magnitude showed the lowering of *M*; these combinations of *JmL* and *JmR* represents FMAFMC category of MSDs. The appearance of *M* lowering at the two opposite corner of Figure 3B signifies that *JmL* and *JmR* have to have opposite sign. This transition in *M* is consistent with the lowering of *M* with *kT* for FMAFMC case (Figure 2J). It is clear that availability of sufficient thermal energy is indispensable for molecules to display their effect. At higher *kT* even lower magnitude of *JmL* and *JmR* caused the significant lowering of *M* of the MSD (Figure 3C-D). Further increase in *kT* brought the randomness in two FM electrodes and hence made data noisy. For *kT*=1.6, most of the MSD with a wide range of *JmL* and *JmR* settled in near zero magnetization state. Relatively, higher values of *JmL* and *JmR* for BFMC and BAFMC cases demonstrated relatively higher magnetization as compared to other combinations (Figure 3F).

Heat capacity (*c*) of MSD provided insight about the phase transitions as a function of variation in exchange coupling of molecule with the two FM electrodes. The *c* for different combinations of *JmL* and *JmR* remained significantly low below *kT*<0.8 (Figure 4A). At kT=0.6, *c* was relatively higher for the disparate magnitude of *JmL* and *JmR* or the non-diagonal positions . For *kT*= 0.8 shows that for the opposite, yet higher magnitude of *JmR* and *JmR*, *c* changed significantly (Figure 4B). In fact, this trend matches with the trend of change in *M* for the similar region of *JmL* and *JmR* at *kT* = 0.8 (Figure 3B). As $kT$ increased- lower magnitude *JmL* and *JmR* (but with opposite signs) started showing sharp changes both in *M* and *c* (Figure 3 and 4 C-D). It is noteworthy that these changes are mainly due to the emergence of a new phase in which left and right electrodes' *M* align in the opposite direction with respect to each other ( Figure 2C-F). In this state *M*, i.e. *M=ML+MR*, stay close to zero as equal and opposite magnetization of the two FM electrodes cancel out each other. For the higher temperature, i. e. $kT \geq 1.4$, *c* of the overall device increase significantly as thermal energy is absorbed for damaging the exchange coupling dependent bonds between the nearest neighbors to create disorder in the electrodes. A notable point is seen where the magnitude of *M* remains close to zero due to oppositely aligned electrodes and thermal disorder.

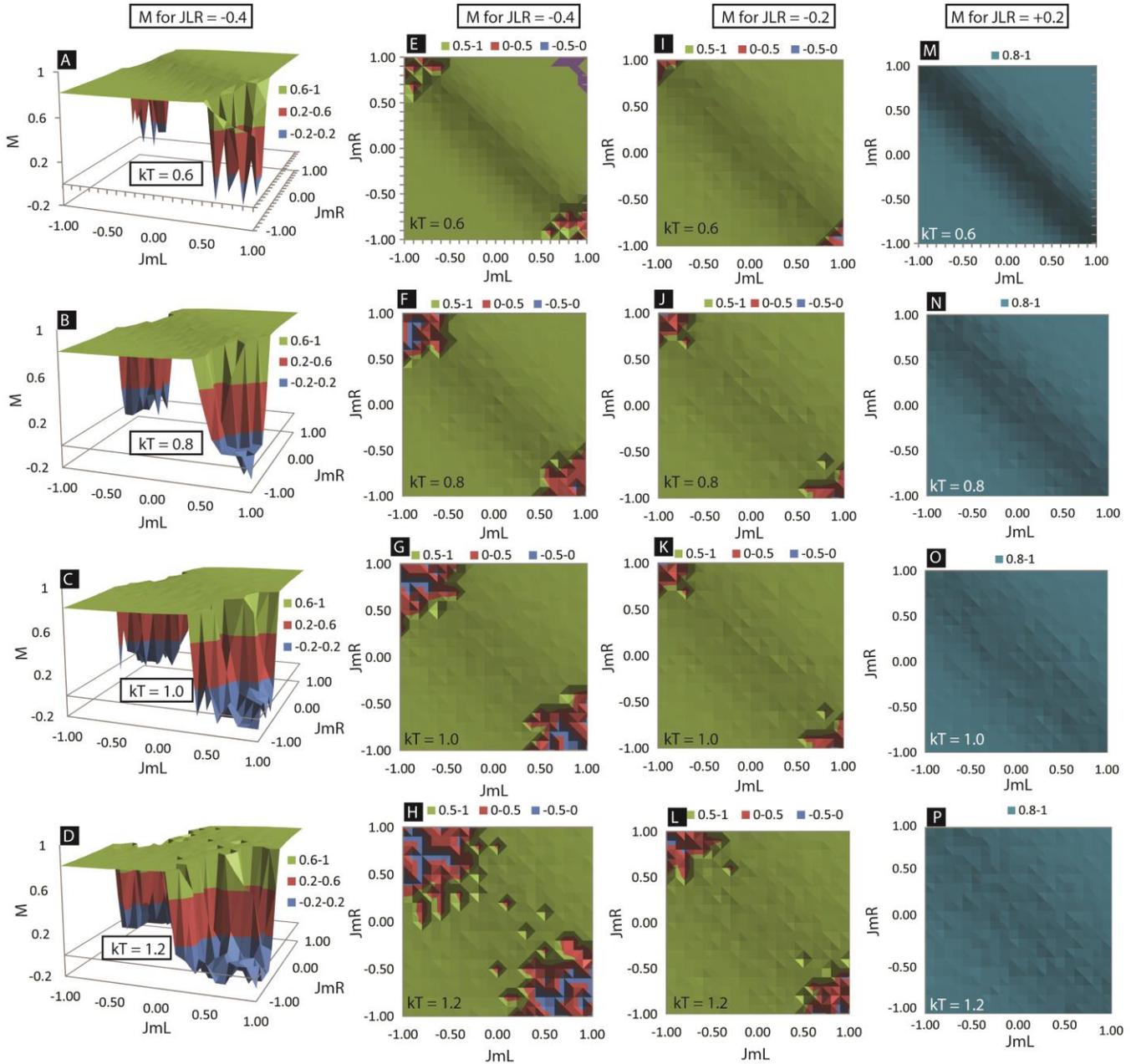

Figure 5: Effect of *JmL* and *JmR* on *M* at different *kT* and *JLR*: 3D graphs for *JL*=-0.4 at *kT* = (A) 0.6, (B) 0.8, (C) 1.0, and (D) 1.2. 2D graphs at *JLR*=-0.4 and *kT*=, (E) 0.6, and (F) 0.8, (G)1.0, and (H) 1.2. 2D graphs at *JLR*=-0.2 and *kT*= (I) 0.6, and (J) 0.8, (K)1.0, and (L) 1.2. 2D graphs at *JLR*=0.2 and *kT*=, (M) 0.6, (N) 0.8, (O)1.0, and (P) 1.2.

Effect of preexisting inter-electrode exchange coupling (*JLR*) was studied on MSD properties. In a real MSD two electrodes can have direct coupling in addition to the coupling via molecule. Strength of direct exchange coupling (*JLR*) between two FM electrodes via space will depend on the inter-electrode gap and the presence of entities, such as impurities, defects, defused atoms in the space region. In reality, sometimes the effect of impurities and defects is much similar to the effect of molecules [10]; in one study on break junction based molecular electronic devices, atomic defects produced Kondo resonance peaks as produced by the molecular device elements. Experimentally, it is very difficult to comprehend various possibilities of deciphering molecule effects on MSD while inter-electrode exchange coupling due to any other reasons, such as small inter-electrode gap, defects and impurities within gap etc., is prevalent. In the context of MSD we have studied the effect of molecule enhanced exchange coupling in the presence of direct inter-electrode coupling (*JLR*).

*JLR* can be both positive (ferromagnetic coupling) and negative (antiferromagnetic coupling). However, in our studies negative *JLR* produced results of crucial significance. Negative *JLR* promoted the emergence of phase change at relatively lower *kT*. We studied the effect of *JLR* for the different combinations of molecular exchange coupling with the left and right FM electrodes, shown by *JmL* and *JmR*, respectively. These studies were performed at various thermal energies. Figure 5(A-D) shows the variation of *M* for 2D MSD of 11x10 size for *JLR* = -0.4. Figure 5 (E-H) shows the top views of the corresponding 3D graphs; these graphs are helpful in monitoring the magnitude of *JmL* and *JmR* required to produce significant changes in the MSDs. Figure 5A and E suggest that with preexisting antiferromagnetic coupling (*JLR*= -0.4) the higher magnitude of oppositely signed *JmL* and *JmR* produced first transition in *M* before *kT* =0.6. For *JLR* = 0, this transition only occurred around *kT* =0.8 (Figure 2F-L and Figure 3B). As temperature increases weaker *JmL* and *JmR* tend to produce phase transition in a MSD for the oppositely signed *JmL* and *JmR*, which is consistent with the results shown in Figure 3 for *JLR* =0. However, the key difference is that the magnitudes of *JmL* and *JmR* for phase transitions is significantly lower at *JLR*= -0.4. For instance at *kT*=0.8 the phase transitions occurred around *JmL* or *JmR* magnitudes to ~0.4, and 0.9 for the *JLR*= -0.4 and *JLR* = 0, respectively. According to these studies, direct inter-electrode antiferromagnetic exchange coupling is a strong factor in making weaker molecular coupling to show the effect. The influence of *JLR* is dependent on its magnitude; comparison of data for *JLR* =-0.4 (Figure 5E-H) and *JLR* =-0.2 (Figure 5I-L) clearly evidenced it. For *JLR* =-0.2 phase transitions occurred at higher magnitudes of *JmL*, *JmR*, and *kT*. Interestingly, negative *JLR* did not affect the areas where both JmL and JmR were positive or negative. Direct antiferromagnetic coupling (-*JLR*) promoted a phase transition in the FMAFMC case, and apparently assisted molecule induced coupling.

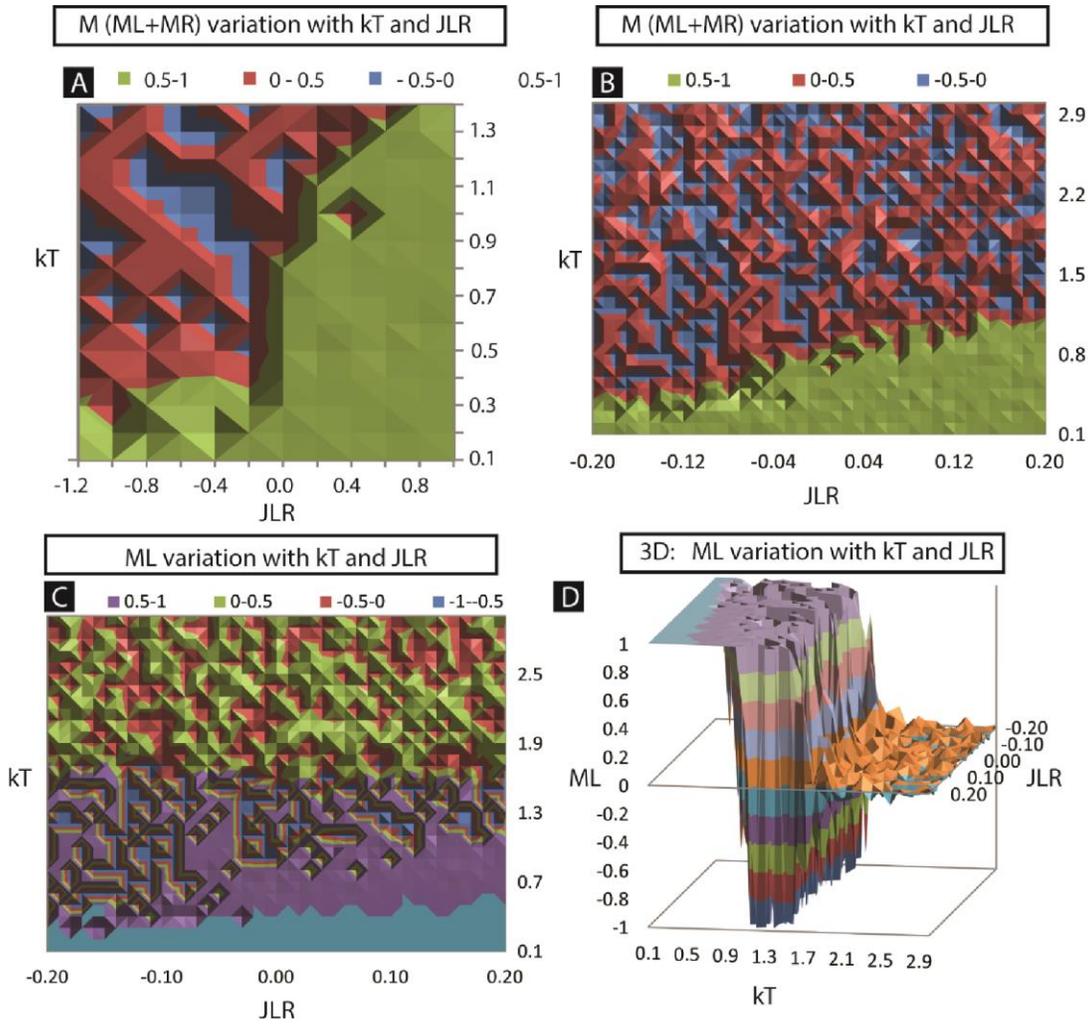

Figure 6: Effect of *JLR* and transitional *kT* on *M* of FMAFMC-MSD: Variation of *M* as (A) *kT* vs. -1.2 to 1.0 *JLR* range,(B) *kT* vs. -0.2 to 0.2 *JLR* range. (C) 2D and (D) variation of *ML* as a function of *JLR* changes.

Effect of positive *JLR*, direct ferromagnetic coupling between two electrodes, was also investigated. Similar to the case of *-JLR*, the *+JLR* also did not produce or affect the phase transition events for the same sign *JmL* and *JmR in* BFMC and BAFMC cases. However, interestingly +JLR made it harder for molecule to produce phase transition for FMAFMC-MSDs. In this case, phase transition required a higher magnitude of *kT, JmL* and *JmR*. For instance, *JLR*=0.4 could only produce phase transition for *kT*=~1.2 and needed *JmL* and *JmR* magnitude to be ~0.8. As discussed earlier *JLR*=0 and *JLR*=-0.4 produced phase transition at *kT*= ~0.8 and *kT*= ~0.4, respectively. We also studied the effect of *JLR* on *c* and *x*; changes in these observable were consistent with the change in *M*. These results are crucial in designing a MSD for a desired temperature range. The working temperature range of a MSD will depend on *JLR* as well.

Since the *JLR* mainly affected the FMAFMC-MSD hence this system was further studied. This study focused on varying *kT* and *JLR* over a wide range for the molecular devices possessing *JmL*=1 and *JmR*= -1. Figure 6A evidenced that negative *JLR* allowed the transition to happen at much lower *kT*. Interestingly, after *JLR*=~ -0.2 transition point remain significantly unchanged. However, increasing +*JLR* increased the magnitude of *kT* transition point; for instance, for *JLR*=~0.2 the phase transition point was after $kT$=~1.2. Effect of *JLR* was very prominent and sharp in the small range of ±0.2. To explore this range we performed simulations (Figure 6B). This study show significant change in transition point occurred around $kT$=-0.1. It is noteworthy that the magnitude of *JLR* is ~10% of *JmL* and *JmR* but, it reduced the transition point by more than 100%. It must be noted that change in *M* from high to low magnitude represent two major transitions in the FMAFMC-MSD case. Second transition, representing thermal energy induced ordered to disordered form around $kT = 1.5$ is not observable in Figure 6B. Observable M remained around zero after both phase transitions and hence one cannot comment on the occurrence of the second phase transition, which is only apparent from the study of *ML* and *MR*.

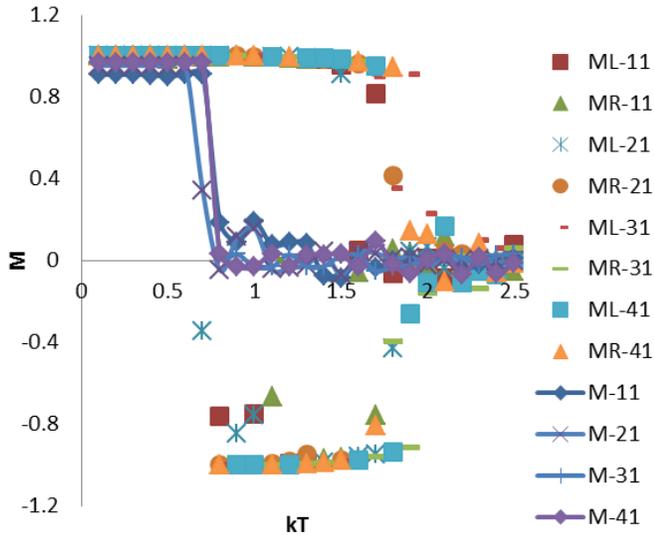

Figure 7: Effect of size variation on *M* and *ML* and *MR* of FMAFMC-MSD.

To study the effect of *JLR* on the second transition we investigated the *ML* and *MR*. For instance, magnetization of the left FM electrode (*ML*) clearly showed the occurrence of the second phase transition in the $kT = 1.5$ range. 3D form of the data shown in Figure 6B provided clear insight about the nature and magnitude of change in the ML (Figure 6D). This study shows the emergence of three phases for the FM electrodes on either side of the molecules; this result is consistent with previously discussed data (Figure 2F-L). In summary, the data for single FM electrode (Figure 6C-D) revealed the second phase transition, which was not observable in the study of total *M* (Figure 6A-B).

Our MCSs also explored the spatial range of molecules' effect. How far away can the effect of the molecule propagate in the ferromagnetic electrode? This question is crucial for deigning the MSD dimension. We studied several 2D and 1D MSDs. Studys on 2D systems with constant height 10 and varying width in 11 to 41 ranges was studied. Up to 41x10 ranges no noticeable change in Curie temperature was observed. For the FMAFMC case no significant change in the *kT* value for the first transition (all electrode in one direction to two electrode in opposite direction) and the second transition (from opposite electrode orientation to complete disordered phase) remained statistically the same (Figure 7). We find it increasingly cumbersome to conduct simulation on very large systems with the desktop computers utilized for these studies. Our 1D simulation on FMAFMC-MSD showed that molecules could reinforce their effect up to several hundred atoms away from them. It means, only few molecular junctions may be sufficient to influence a large area of FM electrodes. In the future we plan to systematically perform 2D simulation with large systems to observe the distance up to which molecules effect can propagate into the FM electrodes.

**CONCLUSION**

Monte Carlo simulations (MCS) were performed to study the effect of magnetic molecule induced exchange coupling on the magnetic properties of the molecular spin devices. We considered all the possible interactions between a magnetic molecule and the two FM electrodes of a MSD. In this study we mainly focused on the Heisenberg type magnetic interaction among nearest neighbors. In BFMC-MSD and BAFMC-MSD cases molecules did not produce much dramatic effect with respect to magnetic properties of the individual electrodes. However, in the FMAFMC-MSD a new phase appeared in the MSD. In this new phase molecules induced the magnetization of the two electrodes in the opposite direction. As a result overall MSD's magnetization ($M= ML+MR$) became zero even when $ML$ and $MR$ were fully ordered. MSD with no direct interaction between FM electrodes exhibited the emergence and termination of this new phase around $kT = $ ~0.8 and $kT=$ ~1.5, respectively. Interestingly, the presence of direct antiferromagnetic coupling dramatically affected the emergence of the new phase. Moreover, molecules' Heisenberg exchange strength was also significantly less for the preexisting antiferromagnetic coupling between two FM electrodes. Effect of molecules was able to penetrate deep into FM electrodes and affected the magnetization of the ferromagnetic electrodes. These MCS are in agreement with the experimental studies of magnetic molecule induced strong coupling between two FM electrodes having direct antiferromagnetic coupling [14]. This study further emphasized that comprehensive design of a successful MSD will involve a careful consideration of direct inter-electrode coupling, FM electrode size, molecule positions, and the strength of interaction between molecules and the electrodes. These parameters will define the basic nature of a molecular spintronics device over a temperature range. In future simulations we will focus on larger molecular spintronics devices incorporating dipolar coupling and anisotropic energy factor of the FM electrodes and biquadratic coupling interactions between molecules and the FM electrodes.

## ACKNOWLDGEMENT

This study was supported by National Science Foundation-Research Initiation Award (Contract # HRD-1238802) and Department of Energy/ National Nuclear Security Agency subaward (Subaward No. 0007701-1000043016). Any opinions, findings, and conclusions or recommendations expressed in this material are those of the author(s) and do not necessarily reflect the views of the National Science Foundation and Department of Energy/National Nuclear Security Administration. Authors thank Gebretesae Tzadu and Ha Huang for assistance with computers set up.